\documentclass[aps,prl,twocolumn,superscriptaddress,floatfix,showpacs]{revtex4}
\usepackage{graphics}
\begin{document}
\title{Supersonic strain front driven by a dense electron-hole
plasma}
\date{\today}
\author{M. F. DeCamp}
\author{D. A. Reis}
\author{A. Cavalieri}
\author{P. H. Bucksbaum}
\author{R. Clarke}
\author{R. Merlin}
\author{E. M. Dufresne}
\author{D. A. Arms}
\affiliation{FOCUS Center and Department of Physics, University of
Michigan}
\author{A. M.  Lindenberg}
\author{A. G. MacPhee}
\affiliation{Department of Physics, University of California,
Berkeley}
\author{Z. Chang}
\affiliation{Department of Physics, Kansas State University}
\author{B. Lings}
\author{J. S. Wark}
\affiliation{Department of Physics, Clarendon Laboratory,
University of Oxford, Oxford, OX1 3PU, UK}
\author{S. Fahy}
\affiliation{Physics Department and NMRC, University College Cork, Ireland.}
\begin{abstract}
We study coherent strain in (001) Ge generated by an
ultrafast laser-initiated high density electron-hole plasma.  The
resultant coherent pulse is probed by time-resolved x-ray
diffraction through changes in the anomalous transmission.
The acoustic pulse front is driven by ambipolar diffusion of the electron-hole
plasma and propagates into the crystal at
supersonic speeds.  Simulations of the strain including electron-phonon coupling, modified by carrier diffusion and Auger recombination, are in good agreement with the observed dynamics.

\end{abstract}
\maketitle
Subpicosecond laser-induced electron-hole plasmas in
semiconductors can produce large amplitude lattice strain and
rapid loss of translational order.  These effects have been
studied extensively in ultrafast linear and nonlinear reflectivity
experiments \cite{shank1975, shank1982, Tom1988, Tinten1995,
Tinten1998, Thomsen1984, Thomsen1986, Chigarev2000, Wright2001}
and, more recently, in time-resolved x-ray Bragg scattering
experiments\cite{rose1999, Lindenberg2000, Siders1999,
Cavalleri2000, Cavalleri2001, Tinten2001}. X-ray diffraction has
the advantage that it can provide quantitative structural
information.
 Many of the x-ray experiments\cite{rose1999,
Lindenberg2000, reis2001} have been analyzed using the
thermoelastic model put forward by Thomsen \emph{et al.}\cite{Thomsen1986} in
which the strain is caused by differential thermal expansion.
Deviations from this model are discussed in the work of 
Thomsen  \emph{et al.} and have been seen in x-ray diffraction
\cite{Lindenberg2000, reis2001, DeCamp2001b,Cavalleri2000, Cavalleri2001}.
Cavalleri \emph{et al.}\cite{Cavalleri2000, Cavalleri2001} studied coherent strain
near the thermal melting threshold in Ge and concluded that the strain
is produced over a region which is thick compared to the optical penetration depth due to 
ambipolar diffusion.  
However, their experiment was only sensitive to structural changes in the near surface region.  

In this letter we report on measurements 
of coherent strain generation in Ge following ultrafast
laser-excitation using a bulk sensitive structural probe. 
We use time-resolved ultrafast x-ray
transmission to measure  strain propagation deep within the crystal, providing information about the generation process.
Initially, the strain front advances at speeds greater than the sound
speed. In our experiments, the laser intensity is sufficient to
impulsively generate a dense electron-hole plasma at the crystal
surface, the dynamics of which are governed by ambipolar diffusion
\cite{Young1982} and Auger recombination.  The plasma couples 
to the lattice through the deformation potential.
In order to
probe the resulting coherent acoustic pulse as it travels deep
within the bulk, we utilize the Laue geometry whereby  the x-rays traverse the
full thickness of the crystal, emerging on the other side as two
mutually coherent beams\cite{batterman-cole}.  We have recently
shown that a short acoustic pulse can coherently transfer energy
between these two beams on a time scale inconsistent with the
thermo-elastic model. Following the initial transient, the beam
intensities oscillate as a function of the
pump-probe delay\cite{DeCamp2001b}.   In the experiments reported
here,  the strain generation is studied as a
function of the incident laser fluence.  The relative phase of the
oscillations and the amplitude of the transient provide
information about the strain generation process at times shorter than the  x-ray probe duration.

In the Laue geometry, two linearly independent wave solutions
propagate through the crystal. Transverse to the propagation,
these two solutions are standing waves whose wavelengths are twice
the spacing of the diffracting planes. The solutions are usually labelled
$\alpha$ and $\beta$ with the convention that $\alpha$ has its
nodes, and $\beta$ its antinodes on the diffracting planes. In the
case that all atoms lie on these planes, $\alpha$ is maximally
transmitted and $\beta$ is maximally absorbed. Because the two
solutions interact with different electron densities, they
propagate with different velocities.

Outside the crystal, two diffracted beams are produced: one in the
direction of the input beam (forward-diffracted or ``0'' beam),
and the other in the direction determined by the vector
sum $\vec{k}_H=\vec{k}_0 +\vec{G}_H$ (deflected-diffracted or
``H'' beam). Here $\vec{k}_{0}$($\vec{k}_{H}$) corresponds to the
wavevector of the forward-diffracted (deflected-diffracted)
beam and $\vec{G}_H$ is the reciprocal lattice vector
corresponding to the diffracting planes. These beams are linear
combinations of the two \emph{internal} solutions, $\alpha$ and
$\beta$.  
The external intensities are given by:
\begin{eqnarray}
\label{fieldeqn1}
I_{0}&=&|a\vec{E}_{\alpha}e^{i\vec{k}_{\alpha}\cdot\vec{z}}+b\vec{E}_{\beta}
e^{i\vec{k}_{\beta}\cdot\vec{z}}|^{2}\\
\label{fieldeqn2}
I_{H}&=&|c\vec{E}_{\alpha}e^{i\vec{k}_{\alpha}\cdot\vec
{z}}-d\vec{E}_{\beta}e^{i\vec{k}_{\beta}\cdot\vec{z}}|^{2}
\end{eqnarray}
where $I_{0}$($I_{H}$) is the diffracted intensity of the forward (deflected)
beam, $\vec{E}_{\alpha,\beta}$ is the
complex wave field inside the crystal, $\vec{k}_{\alpha,\beta}$ is
the complex wavevector of the $\alpha,\beta$ solutions (including absorption), and
$a,b,c,d$ are determined by the crystal orientation. 
The two internal modes $\vec{E}_{\alpha,\beta}$ oscillate in and out of phase 
as they
propagate through the crystal.
The wavelength of the interference, $\Lambda =
|\vec{k}_\alpha-\vec{k}_\beta|^{-1}$, is known as the
Pendell\"osung length which is typically a
few to tens of microns and is often shorter than the absorption
length.

\begin{figure}[tb]
   \includegraphics{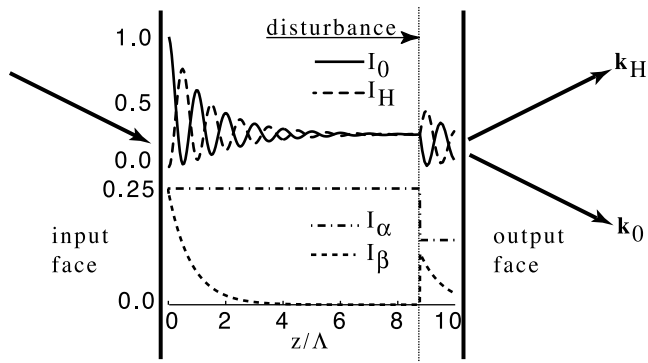}
    \caption{Intensities of the forward- and deflected-diffracted beams (upper) and
    the interior solutions \protect$\alpha, \beta\protect$ (lower) 
    as a function of depth inside a thick crystal.
    A lattice disturbance near the exit couples the 
    two solutions, regenerating \protect$\beta\protect$ .
    } \label{f:regeneration}
\end{figure}

For a crystal that is thick compared to the $\beta$-absorption
length, only the $\alpha$ solution survives and there are no
interference effects. This is the anomalous transmission of x-rays,
known as the Borrmann Effect \cite{batterman-cole}.  A
distortion of the lattice can cause a redistribution of the
interior wave solutions \cite{authier}. Figure
\ref{f:regeneration} shows the effect of a thin region of
distortion regenerating the $\beta$ solution after it has decayed
away in a thick crystal for the case of zero $\alpha$ absorption.
 When this occurs close enough to the
crystal exit, the regenerated $\beta$ wave does not decay away and
interference  occurs at the exit face, despite the fact
that the crystal is thick.

In our experiments, a short acoustic pulse is generated at the
surface of a thick crystal. This pulse can be considered as a
moving lattice disturbance. The diffracted intensities will
oscillate in time as the pulse travels into the crystal bulk with
a period that is given by the Pendell\"osung length divided by the
speed of sound. Deviations from the impulsive strain generation
will be evident in the phase and/or amplitude of the x-ray
modulation as a function of pump-probe delay.
\begin{figure}[tb]
    \includegraphics{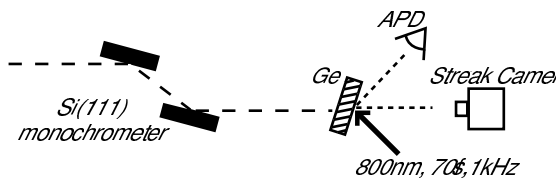} 
    \caption{Experimental setup.} \label{f:setup}
\end{figure}

The experiments were performed at the 7-ID undulator beamline at the Advanced Photon Source. The
x-ray energy was set to 10keV using a cryogenically cooled Si 111
double crystal monochromator leading to a 1.4$\times$10$^{-4}$
fractional energy spread.  The x-ray beam is masked by tantalum
slits to ensure that the x-ray spot is smaller than the laser spot
on the sample and to provide x-ray collimation. The sample is a
280$\mu$m thick, (001) Ge single crystal.  The crystal was
oriented such that the x-rays diffracted in the asymmetric
$20\bar{2}$  Laue geometry. In
this geometry, and at   10 keV, the Pendell\"osung length is 6.2 $\mu$m and the
$\beta$ absorption length is 19$\mu$m, normal to the surface.
Therefore, in the unperturbed crystal, only $\alpha$ survives at
the exit. The only difference between the two diffracted beams is
in their direction and a mismatch in their amplitudes due to
details of the boundary conditions on the exit face
\cite{batterman-cole}.

Coherent strain pulses are produced on the x-ray output face of
the crystal by sub-100fs, 800 nm laser pulses at a 1kHz repetition
rate.  The excitation is fully reversible between subsequent laser
pulses.  The laser is phase-locked to the x-ray beam to better
than the x-ray pulse duration. The laser is timed to the x-rays
using a combination of a digital delay generator and an electronic
phase shifter in the phase-locked loop.  In this manner the
pump-probe delay may be set across a range of $\pm$1 ms with 19 ps
precision. A fast silicon avalanche photodiode (APD) and a
picosecond x-ray streak camera \cite{Chang1996} were used as the
time-resolved detectors. The APD sampled the deflected-diffracted
beam intensity and the streak camera sampled the
forward-diffracted beam (see Fig. \ref{f:setup}). The x-ray
bunch separation was $\sim$152 ns, large enough to allow
electronic gating and measurement of a single x-ray pulse.

\begin{figure}[tb]
    \flushleft \includegraphics{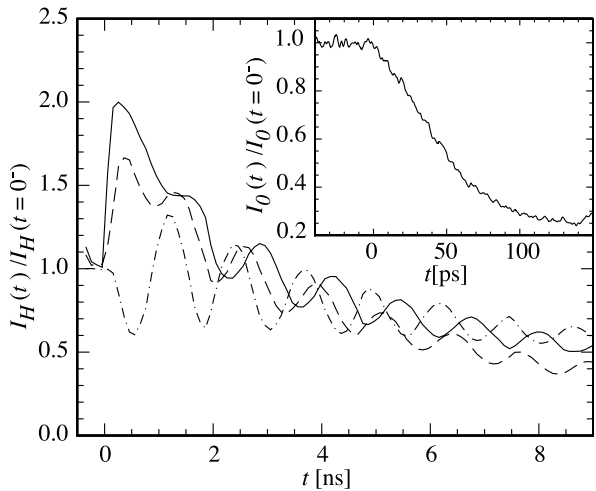}
    \caption{Time-resolved anomalous transmission. The time-dependent
intensity of the deflected-diffracted beam at three different
incident optical fluences: 35
$\frac{\mathrm{mJ}}{\mathrm{cm}^{2}}$ (solid line),
7$\frac{\mathrm{mJ}}{\mathrm{cm}^{2}}$ (dashed line),
2$\frac{\mathrm{mJ}}{\mathrm{cm}^{2}}$ (dot-dashed line).  Inset:
Streak camera data showing the intensity of the forward-diffracted
beam  with picosecond resolution at an incident optical fluence of
35 $\frac{\mathrm{mJ}}{\mathrm{cm}^{2}}$.} \label{f:oscillations}
\end{figure}

\begin{figure}[tb]
   \flushleft \includegraphics{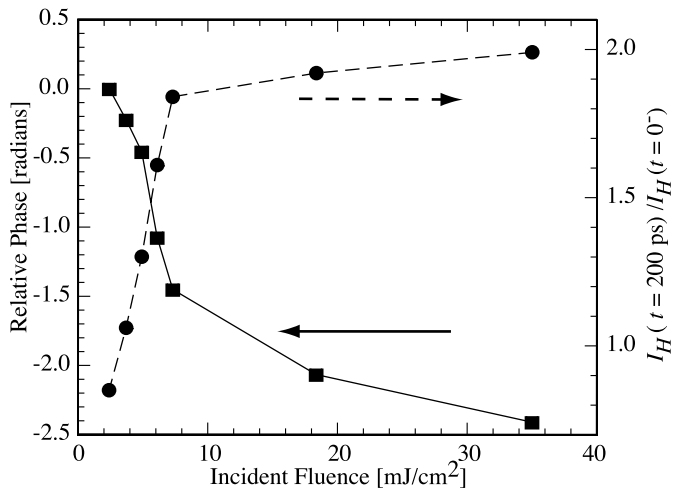}
    \caption{The relative phase of the Pendell\"osung oscillations (squares) and  the normalized deflected-diffracted intensity at a time delay of 200ps  (circles) as a function of incident optical fluence.}     \label{f:phaseshift}
\end{figure}

Following laser-excitation, high contrast oscillations are observed
in the pump-probe data over a large span of excitation densities.
Figure \ref{f:oscillations} shows these oscillations in the
deflected-diffracted beam.  The period of oscillation agrees with 
 the Pendell\"osung length divided by the longitudinal speed of
sound.  At an incident fluence of
35$\frac{\mathrm{mJ}}{\mathrm{cm}^2}$, the behavior near $t=0$ shows
a large transient that is unresolved with the 100 ps x-ray probe
beam.   After the transient, the oscillations show a significant
phase-shift with respect to oscillations that occur following an
excitation of 2$\frac{\mathrm{mJ}}{\mathrm{cm}^2}$. The amplitude
and frequency of the oscillations are relatively insensitive to the
fluence. However, as shown in Fig. \ref{f:phaseshift}, the phase 
is strongly dependent on the fluence
and is correlated with the amplitude of the initial transient.  The relative phase of the
oscillation was defined with respect to  the
2$\frac{\mathrm{mJ}}{\mathrm{cm}^2}$ excitation and was retrieved
from a least squares fit \cite{leastsqr1, leastsqr2}. The
amplitude of the transient is defined as the diffracted intensity
at a delay of 200 ps.  Most of the energy transfer occurs in
$\sim$40ps, measured with the forward diffracted beam using a
streak camera (see the inset in fig.\ref{f:oscillations}). At
relatively high fluences ($>$
10$\frac{\mathrm{mJ}}{\mathrm{cm}^2}$), the intensity of the
deflected diffracted beam approximately doubles while the forward
diffracted beam is cut by more than 75\%. At relatively low
fluences ($<$ 2$\frac{\mathrm{mJ}}{\mathrm{cm}^2}$), there is no
transient.


Inspection of (\ref{fieldeqn1}) and (\ref{fieldeqn2}) shows
that the maximum energy transfer between the forward and deflected
beams near the exit of a thick crystal occurs if the $\alpha$ and
$\beta$ solutions are coupled at a depth of $\Lambda/4$. This
implies that the transient behavior is due to a perturbation to
the lattice that reaches a depth of more than 1.5 $\mu$m into the
bulk. In the simplified picture that a moving interface couples
the $\alpha$ and $\beta$ solutions, the excitation must propagate
into the bulk at greater than 37,000 m/s, more than seven times
the longitudinal speed of sound\cite{footnote}.

The strain pulse has a finite spatial extent and is comprised of a
spectrum of phonons with different wavevectors.  We expect that
the phonon component with wavelength equal to the Pendell\"osung
length will resonantly couple the two interior wave solutions
\cite{Entin1978}. To model this phenomenon, we solve the equations
of dynamical diffraction within the crystal, taking into account
the laser-induced time-dependent strain profiles, using the
Takagi-Taupin formalism adapted for Laue geometry \cite{takagi1962, Taupin1968}. 
In this method,
the differential equations coupling the $\alpha$ and $\beta$
branches are solved numerically.  The depth-dependent strain
profile for a given time is taken
into account by noting that local strain is equivalent to a change
in the local Bragg angle.  Details of this approach (for Bragg
geometry) can be found in the original work of
Takagi\cite{takagi1962} and Taupin \cite{Taupin1968}.  The means
by which the method can be adapted for Laue geometry are implicit
in the work of Zachariasen \cite{zachariasen} and Batterman and
Cole \cite{batterman-cole}.

Pure thermoelastic models of strain propagation do not predict the 
observed fluence dependence of the  phase and amplitude of the Pendell\"osung oscillations.
A proper model must include the effects of the coupling of the photoexcited
plasma to the crystal lattice. 
The strain is comprised of both diffusive and elastic
components:  the diffusive strain is determined by the instantaneous temperature and
carrier density profiles, modified by Auger
recombination; the elastic strain is driven by changes in the temperature
and carrier density, and propagates into the crystal at the speed of
sound. In the absence
of diffusion, a bipolar pulse develops in the time given by the
optical penetration depth divided by the speed of
sound\cite{Thomsen1986}. For LA phonons with wavectors along [100] and a 0.2 $\mu$m
penetration depth, this corresponds to $\sim$40ps. Including
diffusion, the electron-hole plasma extends $\sim$1~$\mu$m in the
same time, leading to a strain front that has propagated into the
bulk faster than the speed of sound.

\begin{figure}[tb]
\flushleft \includegraphics{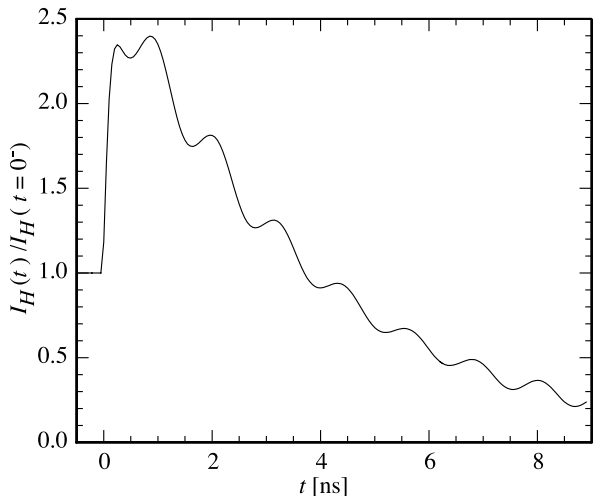} \caption{Simulated
deflected-diffracted intensity for an absorbed laser fluence of
$3\frac{\mathrm{mJ}}{\mathrm{cm}^2}$. }\label{f:ampsim}
\end{figure}

\begin{figure}[tb]
\flushleft \includegraphics{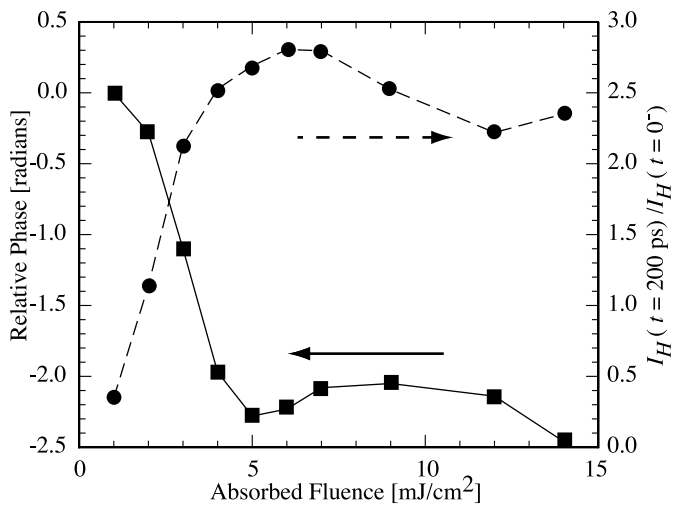} 
\caption{
The calculated relative phase of the Pendell\"osung oscillations (squares) and  the calculated normalized deflected-diffracted intensity at a time delay of 200ps  (circles) as a function of absorbed  optical fluence.}\label{f:phasesim}
\end{figure}


Figure \ref{f:ampsim} shows the calculated diffraction intensity
as a function of laser delay at an \emph{absorbed} laser fluence
of $3\frac{\mathrm{mJ}}{\mathrm{cm}^2}$ (corresponding to a carrier density of $\sim6\cdot10^{20}$ cm$^{-3}$).  Good qualitative
agreement with the experiment is seen (Fig. \ref{f:oscillations}). The sharp
initial rise in diffraction intensity is reproduced, as well as the 
frequency and phase of the time-resolved Pendell\"osung oscillations.
Figure \ref{f:phasesim}
shows the calculated phase shift and the diffracted intensity as a
function of absorbed optical fluence.  After taking into account
the surface reflectivity of the sample, good agreement with the experiment is obtained (Fig. \ref{f:phaseshift}).  

In conclusion we have demonstrated a bulk sensitive probe of lattice dynamics using time-resolved x-ray anomalous transmission.  We have observed that electron-phonon coupling modified by carrier diffusion is a dominant mechanism for energy transport in laser-excited Ge.  This work could be  extended to study how the elastic response of the material can modify the electronic transport properties of semiconductors. 

\begin{acknowledgments}
We thank Bernhard Adams, Marcus Hertlein, Don Walko, and Jared
Wahlstrand for technical assistance and stimulating discussions.
We also thank Jin Wang for use of the intensified CCD camera. This
work was conducted at the MHATT-CAT insertion device beamline at
the Advanced Photon Source and was supported in part by the U.S.
Department of Energy, Grants No. DE-FG02-99ER45743 and No.
DE-FG02-00ER15031, by the AFOSR under contract F49620-00-1-0328 through
the MURI program and from the NSF FOCUS physics frontier center.
One of us (SF) acknowledges the financial support
of Science Foundation Ireland. Use of the Advanced Photon Source was supported by the US Department of Energy Basic Energy Sciences, Office of Energy
Research under Contract No. W-31-109-Eng-38.  
\end{acknowledgments}
\bibliography{GePRL}

\begin{thebibliography}{28}
\bibitem[{foo()}]{footnote}
\bibinfo{note}{Modest strains can reduce the Pendell\"osung length by more than
  one order of magnitude due to the change in the diffraction condition or
  crystal structure factor. However, the associated fringe visability due to
  this effect will be less than 5 percent.}
\end{thebibliography}
\end{document}